\documentstyle[12pt]{article}

\newcommand {\nn}    {\nonumber}
\newcommand {\vs}[1]  { \vspace*{#1 cm} }

\newcounter{eq}
\newcounter{sc}

\newcommand {\MPL}  {Mod.Phys.Lett.}
\newcommand {\NP}   {Nucl.Phys.}
\newcommand {\PL}   {Phys.Lett.}
\newcommand {\PR}   {Phys.Rev.}

\newcommand {\PTP}  {Prog.Theor.Phys.}

\def\overleftrightarrow#1{\vbox{\ialign{##\crcr
 $\leftrightarrow$\crcr\noalign{\kern-1pt\nointerlineskip}
 $\hfil\displaystyle{#1}\hfil$\crcr}}}

\newlength{\minitwocolumn}
\begin{document}


\begin{flushright}
EDO-EP-27\\
July, 1999\\
\end{flushright}
\vspace{30pt}

\pagestyle{empty}
\baselineskip15pt

\begin{center}
{\large\bf On Alternate Derivation of Type IIB Matrix Model

 \vskip 1mm
}

\vspace{20mm}

Ichiro Oda
          \footnote{
          E-mail address:\ ioda@edogawa-u.ac.jp. Address till
         August 4, 1999: Universita Degli Studi Di Padova, 
         Dipartimento Di Fisica "Galileo Galilei", Via F. Marzolo, 
         8, 35131 Padova, Italia, 
         e-mail address:\ ICHIRO.ODA@pd.infn.it
                  }
\\
\vspace{10mm}
          Edogawa University,
          474 Komaki, Nagareyama City, Chiba 270-0198, JAPAN \\

\end{center}


\vspace{15mm}
\begin{abstract}
Starting with Green-Schwarz superstring action, we construct a 
type IIB matrix model. We fix the local $\kappa$ symmetry in the 
Killing spinor gauge and then perform the world-sheet duality
transformation. A matrix model obtained from this gauge-fixed action
is shown to be equivalent to the type IIB matrix model constructed by
Ishibashi et al. Our construction does not make use of an analytic
continuation of spinor variable. Moreover, it seems that our 
construction is applicable to that of more general type IIB matrix 
models in a curved background such as $AdS_5 \times S^5$.

\vspace{15mm}

\end{abstract}

\newpage
\pagestyle{plain}
\pagenumbering{arabic}


\rm

Matrix models provide us non-perturbative formulations of M theory
\cite{BFSS} and type IIB superstring theory \cite{IKKT, FKKT, Aoki,
Iso, Yoneya}.
It is quite surprising that such simple supersymmetric quantum 
mechanics theories seem to contain within them a complicated structure 
of supergravity.

Despite much remarkable success of matrix models, we are still far 
from having a complete understanding of several important problems.
In particular, one big mystery is the problem of background
independence \cite{Witten}. The actions of the matrix models
\cite{BFSS, IKKT} include a flat metric in the kinetic and/or
the potential parts, which is not allowed as stressed in a previous 
article \cite{Oda1}. This is because matrix models, which 
describe quantum gravity, should pick up its own space-time
background in a dynamical manner and should not be $\it{a \ priori}$
formulated in a specific fixed background. Thus it is desirable 
to construct a matrix model which is independent of background
metric \cite{Oda2}, but its relation to quantum gravity is not clear 
at present. Another interesting approach would be construction
of a matrix model in an arbitrary curved background. (But we believe
that the background is not completely arbitrary in that the backgorund
fields would have to satisfy the proper on-shell constraints of 
supergravity in order that the action is $\kappa$-symmetric.) 

Recently, in an attempt of constructing a type IIB matrix model on 
$AdS_5 \times S^5$ \cite{Oda3}, we have utilized the Killing spinor
gauge developed in \cite{Kallosh, Tseytlin} as the gauge-fixing
condition of the $\kappa$-symmetry. (See also related works
\cite{Oda4, Mario, Kimura}.)
The aim of the present article is just to apply the method developed 
there to the case of a flat background geometry. 
Indeed, following the method we will obtain the same matrix model as 
found by Ishibashi et al \cite{IKKT}. One advantage of our construction 
is that we do not have to make use of an analytic continuation of spinor 
variable in the path integral as used by Ishibashi et al. 
Instead, we will use the 
Killing spinor gauge and the world-sheet duality symmetry of type IIB 
superstring theory. 
Another advantage is of course that our construction is extensible to
construction of more complicated matrix models \cite{Oda3}.

First of all, let us start with the type IIB Green-Schwarz
superstring action in a flat background in the Nambu-Goto form
\cite{GS}:
\begin{eqnarray}
S = S_1 + S_2 = \int d^2 \sigma ( L_1 + L_2 ),
\label{1}
\end{eqnarray}
where
\begin{eqnarray}
L_1 &=& - \sqrt{- \frac{1}{2} \Sigma^2}, \nn\\
L_2 &=& - i \varepsilon^{ij} \partial_i X^a (\bar{\theta}^1 \Gamma^a
\partial_j \theta^1 - \bar{\theta}^2 \Gamma^a \partial_j \theta^2)
+ \varepsilon^{ij} \bar{\theta}^1 \Gamma^a \partial_i \theta^1 
\bar{\theta}^2 \Gamma^a \partial_j \theta^2 \nn\\
&=& - i \varepsilon^{ij} {\cal K}^{IJ} \bar{\theta}^I \Gamma^a 
\partial_j \theta^J (\partial_i X^a - \frac{1}{2}i 
\bar{\theta}^K \Gamma^a \partial_i \theta^K),
\label{2}
\end{eqnarray}
with  
\begin{eqnarray}
\Sigma^2 &=& \Sigma_{ab} \Sigma^{ab},     \nn\\
\Sigma^{ab} &=& \Pi^a_i \Pi^b_j \varepsilon^{ij},     \nn\\
\Pi^a_i &=& \partial_i X^a - i \bar{\theta}^I \Gamma^a \partial_i 
\theta^I,     \nn\\ 
{\cal K} &=& \sigma_3 = \pmatrix{1  & 0 \cr 0 & -1 \cr }.
\label{3}
\end{eqnarray}
Here  $\theta^I$ are Majorana-Weyl spinors in ten dimensions and
various indices take the following values: $i, j, ...= 0, 1$; 
$a, b, ...= 0, 1, ..., 9$; $I, J, ...= 1, 2$. 
 
It is straightforward to check that this action has the $N=2$ 
supersymmetry \cite{IKKT}
\begin{eqnarray}
\delta_s \theta^I &=& \varepsilon^I, \nn\\
\delta_s X^a &=& i \bar{\varepsilon}^I \Gamma^a \theta^I,
\label{4}
\end{eqnarray}
and the local $\kappa$-symmetry
\begin{eqnarray}
\delta_{\kappa} \theta^I &=& \alpha^I, \nn\\
\delta_{\kappa} X^a &=& i \bar{\theta}^I \Gamma^a \alpha^I,
\label{5}
\end{eqnarray}
where
\begin{eqnarray}
\alpha^1 &=& (1 + \tilde{\Gamma}) \kappa^1, \nn\\
\alpha^2 &=& (1 - \tilde{\Gamma}) \kappa^2, \nn\\
\tilde{\Gamma} &=& -\frac{1}{2 \sqrt{- \frac{1}{2} \Sigma^2}} 
\Sigma_{ab} \Gamma^{ab}, \nn\\
\tilde{\Gamma}^2 &=& +1.
\label{6}
\end{eqnarray}

Now let us fix the $\kappa$-symmetry in the Killing spinor
gauge \cite{Kallosh, Tseytlin}. The flat Minkowskian space-time
has the Poincare group $ISO(1, 9)$ as the isometry group, so
it is natural to consider the following Killing spinor gauge
condition for the case at hand
\begin{eqnarray}
\theta^I_{-} \equiv {\cal P}_{-}^{IJ} \theta^J = 0,
\label{7}
\end{eqnarray}
where the projection operator ${\cal P}_{\pm}^{IJ}$ is defined as
\begin{eqnarray}
{\cal P}_{\pm}^{IJ} = \frac{1}{2} (\delta^{IJ} \pm i \Gamma_{11}
\varepsilon^{IJ}),
\label{8}
\end{eqnarray}
with the chiral matrix $\Gamma_{11} = \Gamma^0 \Gamma^1 ...\Gamma^9$.
(The other simple choice of the projection operator is
${\cal P}_{\pm}^{IJ} = \frac{1}{2} (\delta^{IJ} \pm i \varepsilon^{IJ})$.
It is easy to see that this choice leads to the same result as that
shown in this paper.)
Note that this projection operator satisfies the following
identities:
\begin{eqnarray}
{\cal P}_{\pm}^2 &=& {\cal P}_{\pm}, \nn\\
{\cal P}_{+} {\cal P}_{-} &=& {\cal P}_{-} {\cal P}_{+} = 0, \nn\\
{\cal P}_{\pm} \Gamma^a &=& \Gamma^a {\cal P}_{\mp}.
\label{9}
\end{eqnarray}
As a result, we can show 
\begin{eqnarray}
\bar{\theta}^I \Gamma^a \partial_i \theta^I &=& 2 i \bar{\theta} 
\Gamma^a \partial_i \theta, \nn\\
{\cal K}^{IJ} \bar{\theta}^I \Gamma^a \partial_i \theta^J &=& 0. 
\label{10}
\end{eqnarray}
where we have defined $\theta_{+}^1 = \theta$. The latter equation
in Eq.(\ref{10}), together with Eq.(\ref{2}), leads to the fact that 
the Wess-Zumino term $L_2$ vanishes in the present gauge condition.
Thus, we arrive at the $\kappa$-symmetry gauge-fixed action given by
\begin{eqnarray}
S = -  \int d^2 \sigma \sqrt{- \frac{1}{2} \Sigma^2},
\label{11}
\end{eqnarray}
where
\begin{eqnarray}
\Sigma^2 &=& \Sigma_{ab} \Sigma^{ab},     \nn\\
\Sigma^{ab} &=& \Pi^a_i \Pi^b_j \varepsilon^{ij},     \nn\\
\Pi^a_i &=& \partial_i X^a - 2i \bar{\theta} \Gamma^a \partial_i \theta.
\label{11-2}
\end{eqnarray}

The $\kappa$-symmetry gauge-fixed action (\ref{11}) is still
quartic with respect to the fermionic variable, but it is possible
to reduce it to a much simpler quadratic form by making the 
world-sheet duality transformation as follows \cite{Tseytlin}.
To this aim, let us first rewrite the action (\ref{11}) into the
Polyakov form 
\begin{eqnarray}
S = - \frac{1}{2} \int d^2 \sigma \sqrt{- g} g^{ij}
(\partial_i X^a - 2i \bar{\theta} \Gamma^a \partial_i \theta)
(\partial_j X^a - 2i \bar{\theta} \Gamma^a \partial_j \theta).
\label{11-3}
\end{eqnarray}
Next, we rewrite the action (\ref{11-3}) into the first-order form 
by introducing the Lagrange multiplier $P^a_i$
\begin{eqnarray}
\tilde{S} = - \frac{1}{2} \int d^2 \sigma \sqrt{-g} g^{ij}
\left[ - P^a_i P^a_j + 2 P^a_i (\partial_j X^a - 2i \bar{\theta} 
\Gamma^a \partial_j \theta) \right].
\label{12}
\end{eqnarray}
Actually the integration over $P^a_i$ gives rise to the action
(\ref{11-3}). After integrating out $X^a$ and then solving the resulting
equation in terms of the dual variable $\tilde{X}^a$ as
\begin{eqnarray}
P^{ia} = \frac{1}{\sqrt{-g}} \varepsilon^{ij} \partial_j \tilde{X}^a,
\label{13}
\end{eqnarray}
we can obtain the desired dual action
\begin{eqnarray}
\tilde{S} = - \frac{1}{2} \int d^2 \sigma (\sqrt{-g} g^{ij}
\partial_i \tilde{X}^a \partial_j \tilde{X}^a 
+  4i \varepsilon^{ij} \partial_i \tilde{X}^a \bar{\theta} 
\Gamma^a \partial_j \theta).
\label{14}
\end{eqnarray}
It is quite remarkable that the $\kappa$-symmetry gauge-fixed action 
(\ref{11}) is transformed to a much simpler quadratic form through 
the world-sheet duality transformation. 

The action $\tilde{S}$ is still invariant under the residual $N=2$
supersymmetry. This supersymmetry can be found by amalgamating
the original supersymmetry (\ref{4}) with the $\kappa$-symmetry
(\ref{5}) to preserve the gauge fixing condition (\ref{7}), in
other words, $\theta^1 = i \Gamma_{11} \theta^2$. Following a
similar procedure to the reference \cite{IKKT}, let us consider 
the mixed transformation of the $N=2$ supersymmetry (\ref{4}) 
and the $\kappa$-symmetry (\ref{5})
\begin{eqnarray}
\delta \theta^I &=& \delta_s \theta^I + \delta_{\kappa} 
\theta^I, \nn\\
\delta X^a &=& \delta_s X^a + \delta_{\kappa} X^a.
\label{15}
\end{eqnarray}
Provided that we define $\tilde{\varepsilon}^2 = i \Gamma_{11}
\varepsilon^2$ and $\tilde{\kappa}^2 = i \Gamma_{11} \kappa^2$
and then choose
\begin{eqnarray}
\kappa^1 &=& \frac{- \varepsilon^1 + \tilde{\varepsilon}^2}{2}, 
\nn\\
\tilde{\kappa}^2 &=& \frac{ \varepsilon^1 - \tilde{\varepsilon}^2}{2}, 
\label{16}
\end{eqnarray}
the gauge fixing condition is preserved as desired.
With these conditions and the definition
\begin{eqnarray}
\xi &=& \frac{ \varepsilon^1 + \tilde{\varepsilon}^2}{2}, 
\nn\\
\varepsilon &=& \frac{ \varepsilon^1 - \tilde{\varepsilon}^2}{2}, 
\label{16-2}
\end{eqnarray}
it is easy to derive the new $N=2$ 
supersymmetry law of $\theta$ whose result is given by
\begin{eqnarray}
\delta \theta = \xi - \tilde{\Gamma} \varepsilon,
\label{17}
\end{eqnarray}
whereas it is necessary to make some work to do the transformation 
law of $\tilde{X^a}$. In a similar way to the case of $\theta$, 
we can find the new $N=2$ supersymmetry transformation
law of the original variable $X^a$
\begin{eqnarray}
\delta X^a = - 2i \bar{\theta} \Gamma^a (\xi + \tilde{\Gamma} 
\varepsilon),
\label{18}
\end{eqnarray}
which gives us the transformation of $P^a_i$ 
\begin{eqnarray}
\delta P^a_i = - 4i \partial_i \bar{\theta} \Gamma^a \tilde{\Gamma} 
\varepsilon.
\label{19}
\end{eqnarray}
{}From this expression, we can derive the the new $N=2$ 
supersymmetry law of $\tilde{X^a}$
\begin{eqnarray}
\delta \tilde{X}^a = 4i \bar{\varepsilon} \Gamma^a \theta.
\label{20}
\end{eqnarray}
Note that to derive this equation we have used the fact that 
$g_{ij} \tilde{\Gamma}$ can be replaced with $\frac{\varepsilon_{ij}}
{\sqrt{-g}}$ in $\delta P^a_i$ since both the expressions
make the same contribution in the action $\tilde{S}$.
At this stage, it is important to notice that $\tilde{\Gamma}$ can be
expressed in terms of the dual variable as
\begin{eqnarray}
\tilde{\Gamma} = \frac{1}{2 \sqrt{-g}} \partial_i \tilde{X}^a 
\partial_j \tilde{X}^b \varepsilon^{ij} \Gamma^{ab}.
\label{21}
\end{eqnarray}

{}Finally, integrating out $g_{ij}$, we finish with the dual
action
\begin{eqnarray}
\tilde{S} = - \int d^2 \sigma (\sqrt{- \frac{1}{2} \sigma^2}
+  2i \varepsilon^{ij} \partial_i X^a \bar{\theta} 
\Gamma^a \partial_j \theta).
\label{22}
\end{eqnarray}
The $N=2$ supersymmetry of this action is then given by
\begin{eqnarray}
\delta \theta &=& - \frac{1}{2 \sqrt{-\frac{1}{2} \sigma^2}}
\sigma_{ab} \Gamma^{ab} \varepsilon + \xi, \nn\\
\delta X^a &=& 4i \bar{\varepsilon} \Gamma^a \theta.
\label{23}
\end{eqnarray}
In the above equations, we have written $\tilde{X}^a$ simply as
$X^a$ and defined $\sigma^{ab}$ as
\begin{eqnarray}
\sigma^{ab} = \partial_i X^a \partial_j X^b \varepsilon^{ij}.
\label{24}
\end{eqnarray}
These expressions are equivalent to those in \cite{IKKT}, so
we can obtain the same type IIB matrix model by following
the same path of thought as in \cite{IKKT}. Notice that even if
we have derived the same matrix model by starting with the same
type IIB superstring action (\ref{1}), (\ref{2}) the derivation 
method is different. In particular, our approach does not rely on an
analytic continuation of the spinor variable in the path integral
which plays an important role in \cite{IKKT} and seems to be 
extensible to more general situation \cite{Oda3}.
Of course, the key idea behind the approach at hand is a symmetry of
the Green-Schwarz superstring action \cite{GS} under the world-sheet
duality transformation. Incidentally, we have recently examined
the duality transformations of the Green-Schwarz superstring as
well as supersymmetric D-brane actions in a general type II
supergravity background \cite{Oda5}.

\vs 1

\end{document}